\documentclass[a4paper,11pt]{article}
\usepackage{pos}
\usepackage{graphicx}  % needed for figures
\usepackage{dcolumn}   % needed for some tables
\usepackage{bm}        % for math
\usepackage{standalone}
\usepackage{enumitem}
\usepackage[absolute,overlay]{textpos}
\usepackage{xcolor}
\usepackage{slashed}
\usepackage{booktabs}
\usepackage{multirow}
\usepackage{amsmath}
\usepackage{bbm}
\usepackage{bbold}
\usepackage{stackrel}
\usepackage{rotating}
\usepackage{pifont}
\usepackage{mathtools}
\usepackage{pgfplots}
\usepackage{tikz}
\usetikzlibrary{calc}

\definecolor{myred}{RGB}{227,0,31}
\definecolor{myblue}{RGB}{56,125,216}
\definecolor{mygreen}{RGB}{84,122,58}
\usepackage{hyperref}
\hypersetup{
    colorlinks=true,       % false: boxed links; true: colored links
    linkcolor=blue,          % color of internal links
    citecolor=blue,        % color of links to bibliography
    filecolor=blue,      % color of file links
    urlcolor=blue           % color of external links
}
\usepackage{simplewick}

\usepackage{float} % Forces placement of figures
\usepackage{tikz} % adds /foreach (looping) command
\usepackage{xspace}

\usepackage{subfig}

\title{Exploring the large-$N_c$ limit with one quark flavour}
\ShortTitle{Exploring the large-$N_c$ limit with one quark flavour}

\author[a]{Michele~Della~Morte}
\author*[a,b]{Benjamin~J{\"a}ger}
\author[a,b]{Sofie~Martins}
\author[b,c,d]{Francesco~Sannino}
\author[a,e]{J.~Tobias~Tsang}
\author[f]{Felix~P.~G.~Ziegler}

\affiliation[a]{CP$^3$-Origins, Dept. of Mathematics and Computer Science, University of Southern Denmark, Campusvej 55, 5230 Odense M, Denmark}
\affiliation[b]{Danish Institute for Advanced Study, University of Southern Denmark, Campusvej 55, 5230 Odense M, Denmark}
\affiliation[c]{CP$^3$-Origins, Dept. of Physics, Chemistry and Pharmacy, University of Southern Denmark, Campusvej 55, 5230 Odense M, Denmark}
\affiliation[d]{Dipartimento di Fisica “E. Pancini", Università di Napoli Federico II - INFN sezione di Napoli, Complesso Universitario di Monte S. Angelo Edificio 6, via Cintia, 80126 Napoli, Italy}
\affiliation[e]{Department of Theoretical Physics, CERN, 1211 Geneva 23, Switzerland} 
\affiliation[f]{School of Physics and Astronomy, The University of Edinburgh, EH9 3FD Edinburgh, United Kingdom}

\emailAdd{jaeger@imada.sdu.dk}

\abstract{
We use one-flavour QCD ($N_c=3$) as a proxy to understand $\mathcal{N}=1$ SYM. For our simulations, we use tree-level improved Wilson fermions and a Symanzik improved gauge action. The hadron spectrum is obtained by using LapH smearing for different masses and simulation volumes. We also report on our efforts to increase the number of colours in our simulations, where we find that the simulations show  increasing topological freezing for larger $N_c$.

\begin{textblock}{20}(15.0,1.70)
CERN-TH-2022-209\\
\end{textblock}% 
}

\FullConference{%
The 39th International Symposium on Lattice Field Theory,\\
8th-13th August, 2022,\\
Rheinische Friedrich-Wilhelms-Universität Bonn, Bonn, Germany
}

\begin{document}
\maketitle

\section{Introduction}
We follow the Corrigan and Ramond large-$N_c$ expansion~\cite{Corrigan:1979xf} to check the predictions from super-symmetric Yang-Mills (SYM) theories. The large $N_c$ dynamics of this limit was investigated in~\cite{Sannino:2007yp} with crucial differences with respect to the ordinary large $N_c$ dynamics. Additionally at leading order in this limit, it emerges a connection with the spectrum and dynamics of super Yang Mills~\cite{Armoni:2003gp,Armoni:2003fb,Sannino:2003xe} with nontrivial consequences also for the thermodynamics of the system~\cite{Sannino:2005sk}. Assuming that $N_c=3$ is large, we check whether such predictions are realised by the data. A previous study~\cite{Farchioni:2007dw} computed the mesonic spectrum. We improve upon their findings by considering a finer lattice spacing, larger volumes, and tree-level improved actions. In the near future, we plan to extend this to $N_c=4, 5$ and $6$.

\section{Lattice Setup}

For three numbers of colours ($N_c=3$) the two-index anti-symmetric representation coincides with the conjugate representation, which allows us to employ "standard" lattice QCD simulations and methods. For our lattice simulations we use one flavour of tree-level improved Wilson fermions ($c_{sw} = 1$) at a fixed gauge coupling of $\beta = 4.5$. For the gluonic degrees of freedom, we utilise the Symanzik-improved gauge action~\cite{Iwasaki:1983iya}. We use the openQCD software package~\cite{openQCD} and, since we only simulate a single fermion species, we rely solely on the RHMC algorithm~\cite{Clark:2006fx}. For the approximation needed in the RHMC, we use a Zolotarev functional with degree of tenth order. The integration scheme in the HMC is divided into three levels with two Omelyan 4th and one 2nd order integrators~\cite{OMELYAN2003272}. The number of integration steps is tuned to achieve a high acceptance, which we find to be at least $84\%$ for all our ensembles. To suppress auto-correlation effects we separate configurations by at least 64 MD units. 
\begin{table}
    \centering
    \begin{tabular}{c|cccccccccc}
    \toprule
$L/a$      & 12   & 12   & 12   & 12   & 16    & 16   & 16   & 16   & 16   & 16   \\
$\kappa$  & 0.135 & 0.137 & 0.139 & 0.140 & 0.135  & 0.137 & 0.139 & 0.140 & 0.1405 & 0.141 \\
$N_{\mathrm{cnfg}}$ & 7007 & 6210 & 5840 & 5380 & 12059 & 4285 & 9458 & 8465 & 6184 & 7879 \\
\midrule
$L/a$      & 20   & 20   & 20   & 24   & 24   & 24   & 24   & 32   & 32    & \\ 
$\kappa$   & 0.135 & 0.137 & 0.139 & 0.135 & 0.139 & 0.1405 & 0.141 & 0.139 & 0.140 & \\ 
$N_{\mathrm{cnfg}}$ & 4046 & 1435 & 2747 & 2844 & 2540 & 2242 & 2000 & 1429 & 428  & \\ 
\bottomrule
    \end{tabular}
    \caption{Overview of lattice ensembles.     }
    \label{tab.over}
\end{table}
Table~\ref{tab.over} shows an overview of the ensembles generated for this study, which spans a wide range of volumes and hopping parameters $\kappa$. To compare this setup with other lattice simulations, we apply the Wilson flow~\cite{Luscher:2010iy} to the Yang-Mills action density and obtain an indicative lattice spacing of $a=0.06\,$fm. For this we assumed $\sqrt{8\, t_0} = 0.45\,$fm, which is between the values for $N_f=0$~\cite{Luscher:2010iy} and $N_f=2$~\cite{Bruno:2013gha}.

Simulations of one flavour of Wilson fermions may suffer a sign problem. We discussed
this issue in our previous Lattice contribution~\cite{Ziegler:2021nbl}, where we
presented results for the sign of the Wilson Dirac operator on our ensembles.

\section{One Flavour for $N_c = 3$}
\begin{figure}
   \includegraphics[width=0.98\textwidth]{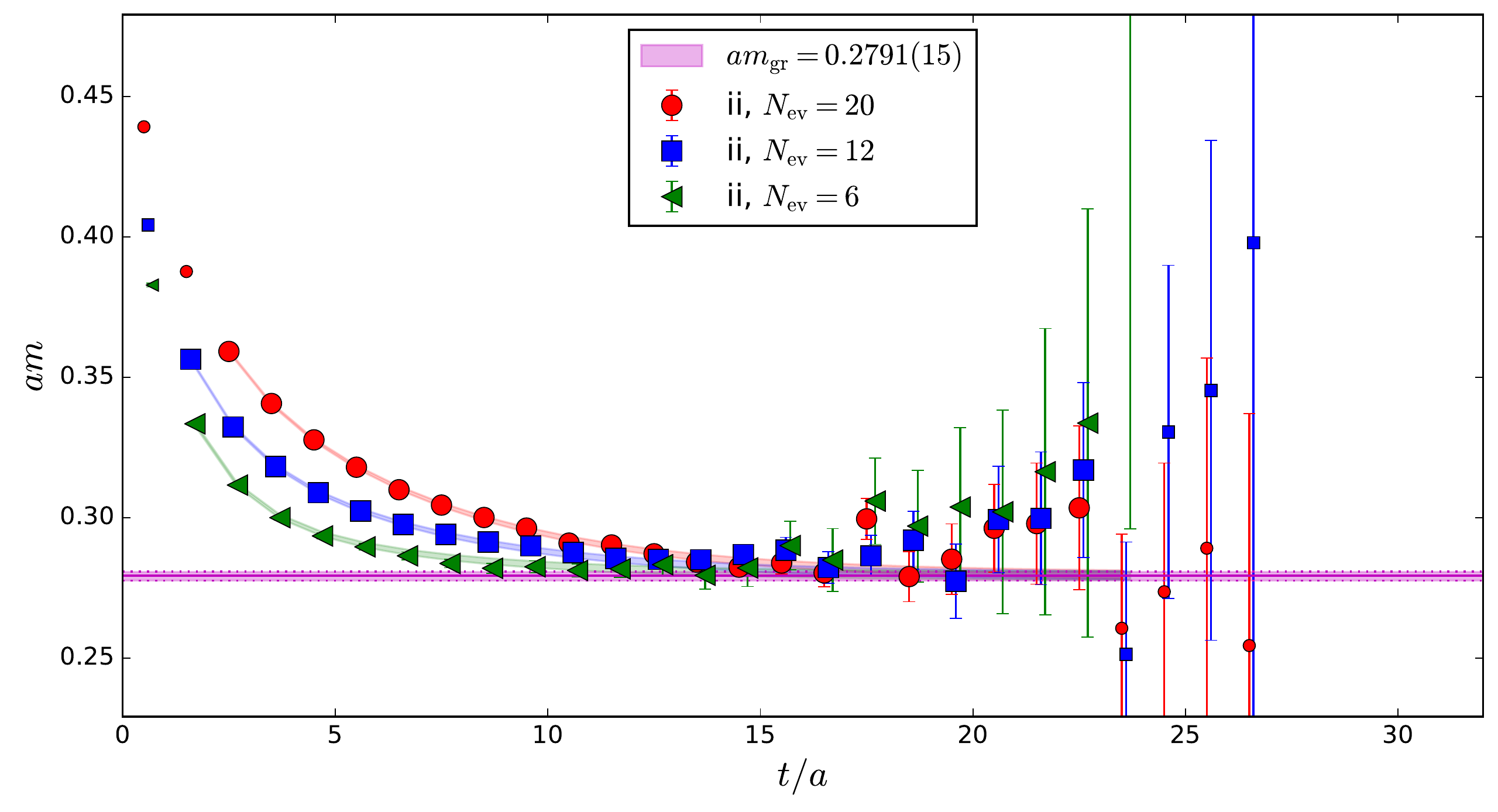}
\caption{Example of a simultaneous fit to three vector-vector correlation functions with different LapH smearings (indicated by the differently coloured markers). The spectrum is  extracted using a three-state fit ansatz.
We show the effective masses for the three correlation functions (data points). We superimpose the ground state fit result (magenta horizontal band) and the approach to the plateau (faintly coloured bands).}
\label{fig.example_fit}
\end{figure}

To check predictions from $\mathcal{N}=1$ SYM we study the spectrum of mesons. Since only a single fermion species is present, disconnected diagrams appear in all correlation functions. To include disconnected diagrams we use the Laplacian Heaviside (LapH) method~\cite{Morningstar:2011ka,Peardon:2009gh}. For comparison, we also compute the purely connected correlation function as done in standard lattice calculations. We use the connected correlation function of the pseudo-scalar to define the chiral point, i.e. where the quark mass vanishes. As the connected pseudo-scalar is identified in standard lattice simulations with the pion, we name the corresponding ground state \emph{fake pion}. This nomenclature is used to emphasise that this state is unphysical in our setup.

The LapH method has the additional advantage that we can change the level of smearing by changing the number of eigenvalues used in the approximation of the Laplacian Heaviside kernel. Due to the multiple smearings, we have access to more data, with the same spectrum but different overlap coefficients. As can be seen from the effective mass plot in Figure~\ref{fig.example_fit}, different smearing choices result in different approaches to the ground state mass.

We perform a combined correlated fit to multiple correlation functions, which differ in their LapH smearings. The fit ansatz includes three states, giving access to the ground state and two excited states. However, we only use the two lower masses in our subsequent analysis as the second excited state is prone to fit systematics from the choice of fitrange.
The bands in Figure~\ref{fig.example_fit} show the effective mass behaviour of the fit result for the correlation functions that enter the fit. The horizontal magenta band indicates the ground state mass obtained from the fit.

\begin{figure}
\centering
\subfloat{
  \includegraphics[width=0.49\textwidth]{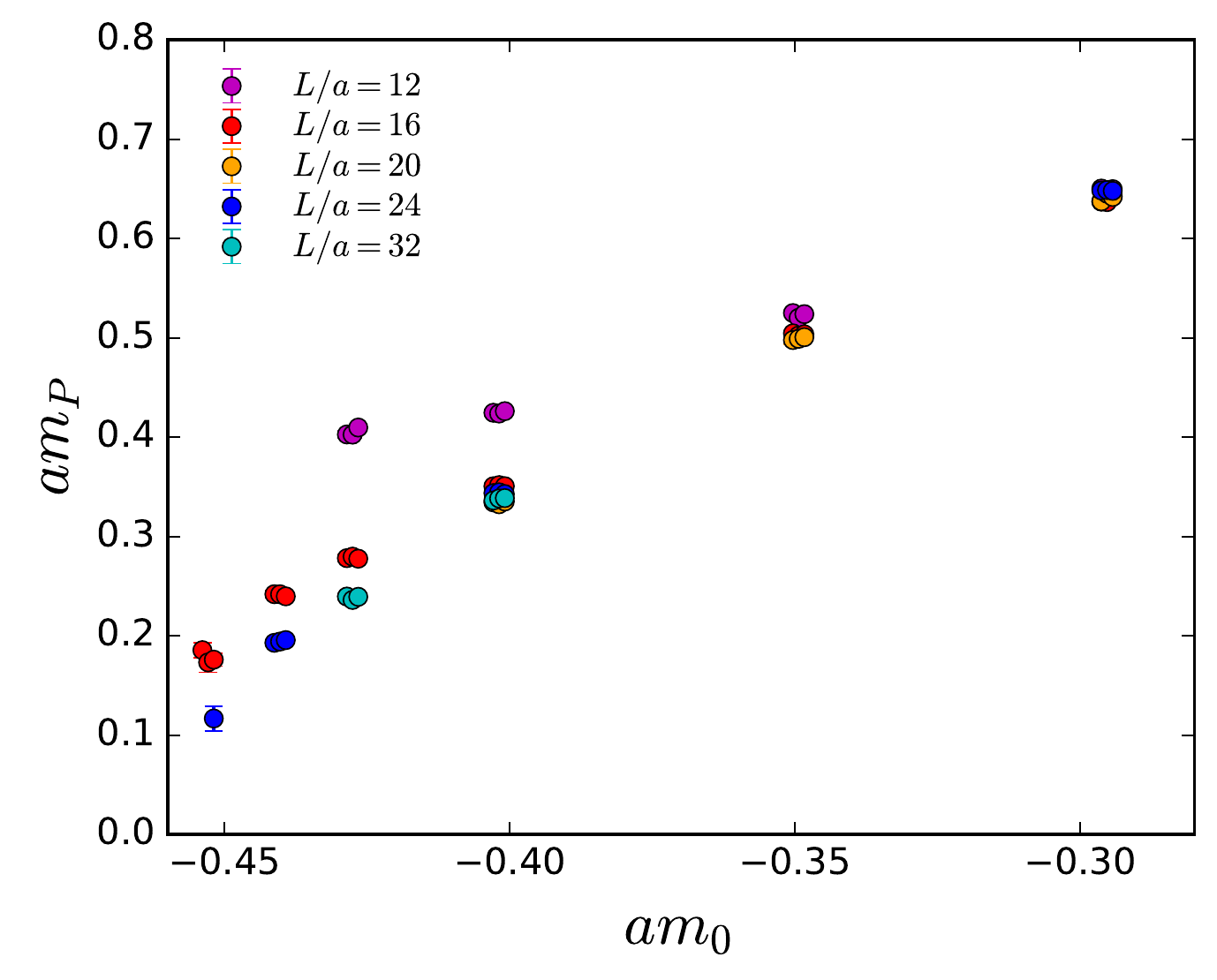}
}
\subfloat{
  \includegraphics[width=0.49\textwidth]{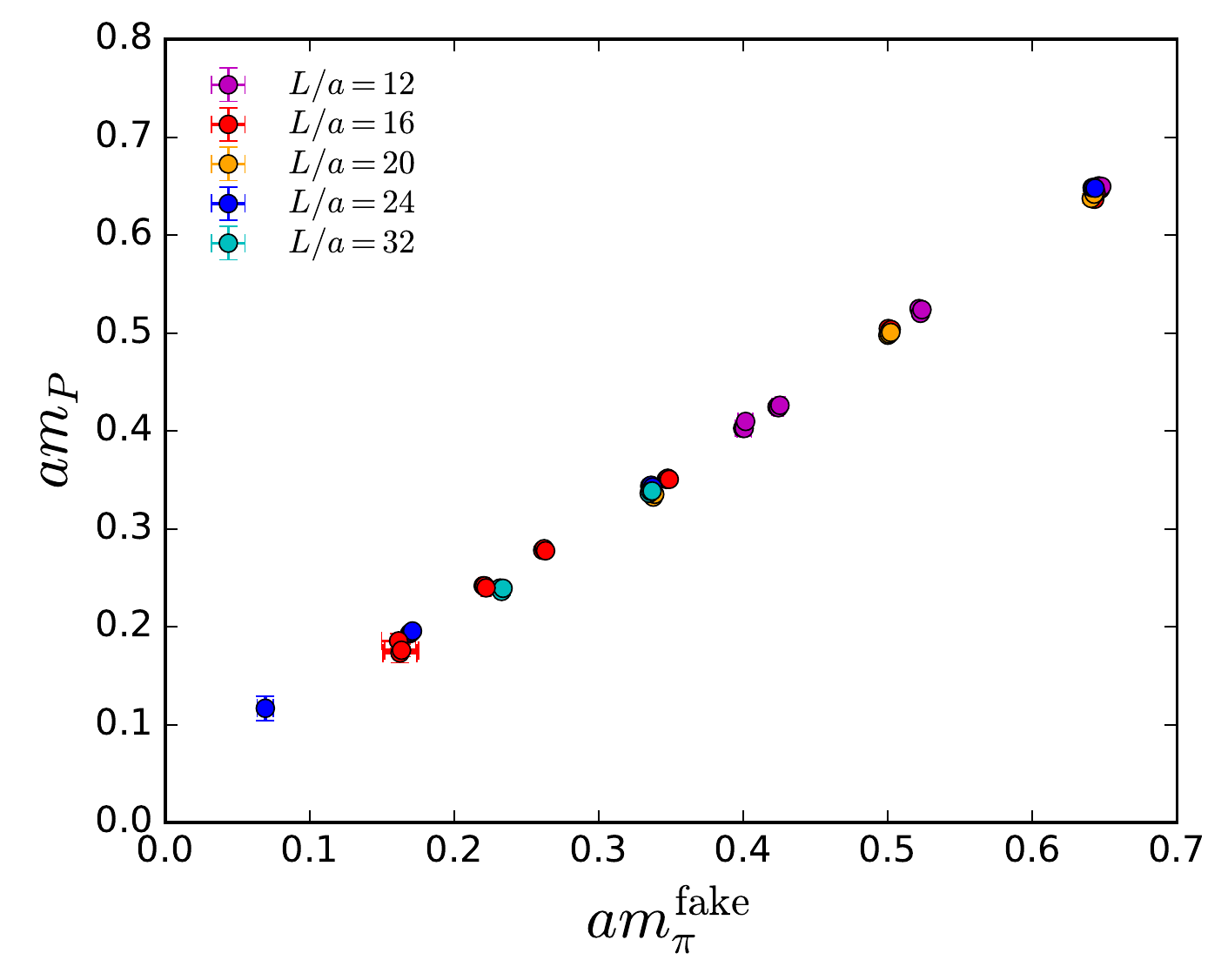}
}
\caption{Results for the pseudo-scalar meson as a function for the bare quark mass (left) and the fake pion mass (right).}
\label{fig.ps_kL}
\end{figure}
Figure~\ref{fig.ps_kL} shows the extracted pseudo-scalar masses as a function of the bare quark mass $m_0 = \tfrac{1}{2 \kappa} - 4$ on the left and of the connected pseudo-scalar mass ($m_\pi^\mathrm{fake}$) on the right. Using the fake pion mass instead of the bare quark mass, we obtain a significantly smoother and less volume-sensitive behaviour. Furthermore, this choice facilitates the definition of the chiral limit, as it eliminates the need to determine the value $\kappa_\mathrm{crit}$ of the hopping parameter at which the quark mass vanishes.

\begin{figure}
\centering
\subfloat{
  \includegraphics[width=0.49\textwidth]{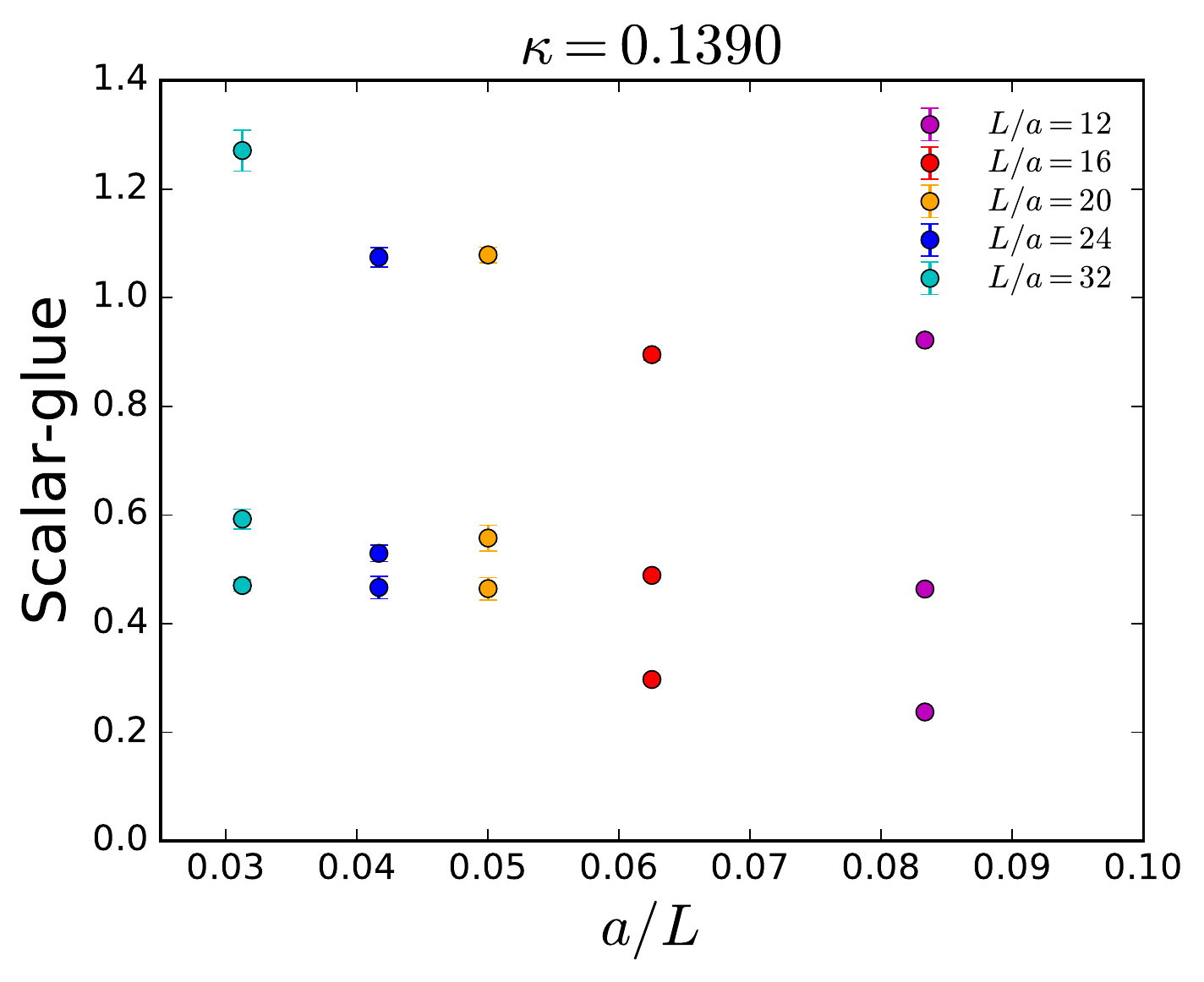}
}
\subfloat{
  \includegraphics[width=0.49\textwidth]{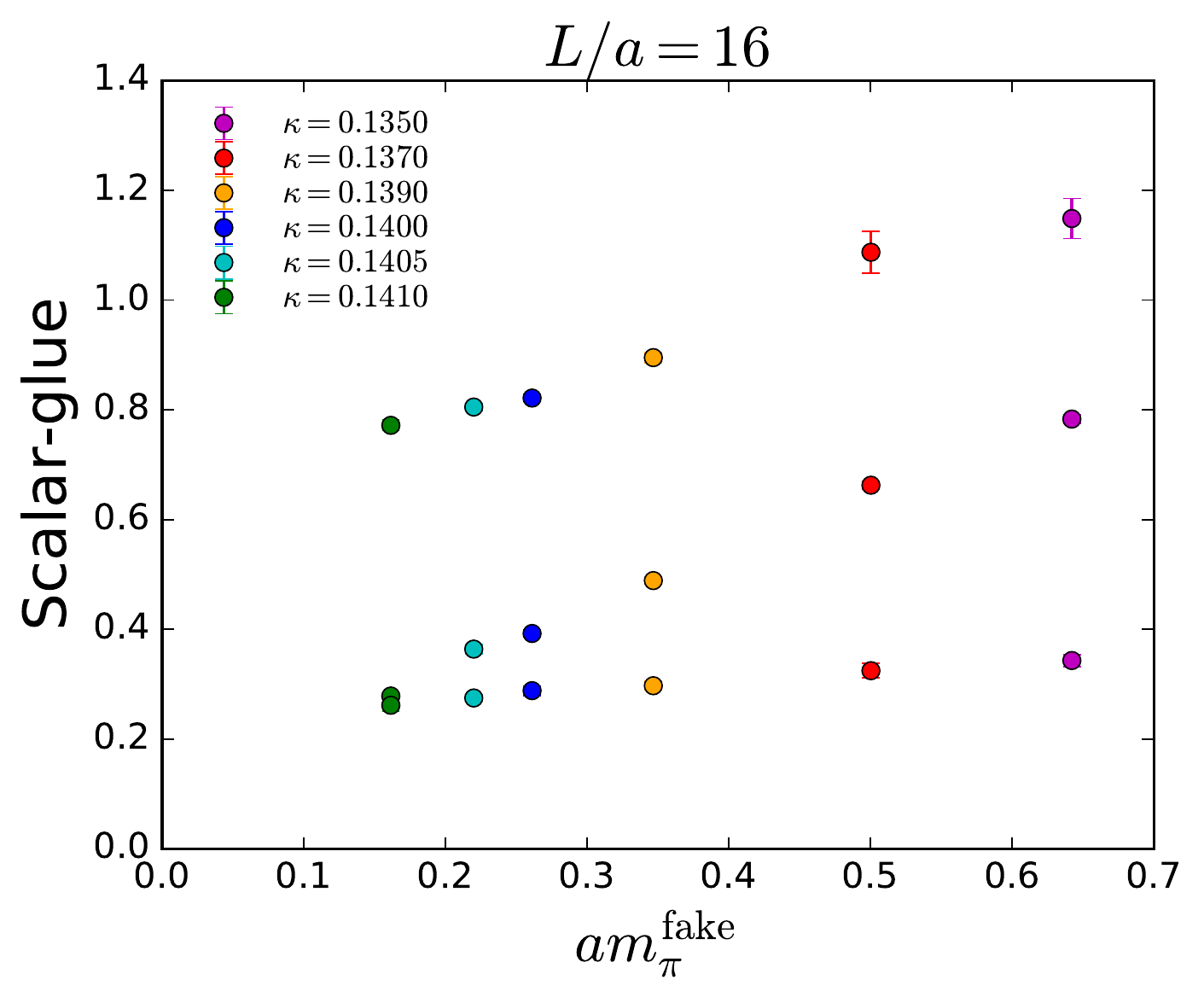}
}
\caption{Results of the scalar meson as a function of volume (left) and the mass of the fake pion (right).}
\label{fig.scalar}
\end{figure}
\begin{figure}
\centering
\subfloat[Pseudo-scalar]{
  \includegraphics[width=0.49\textwidth]{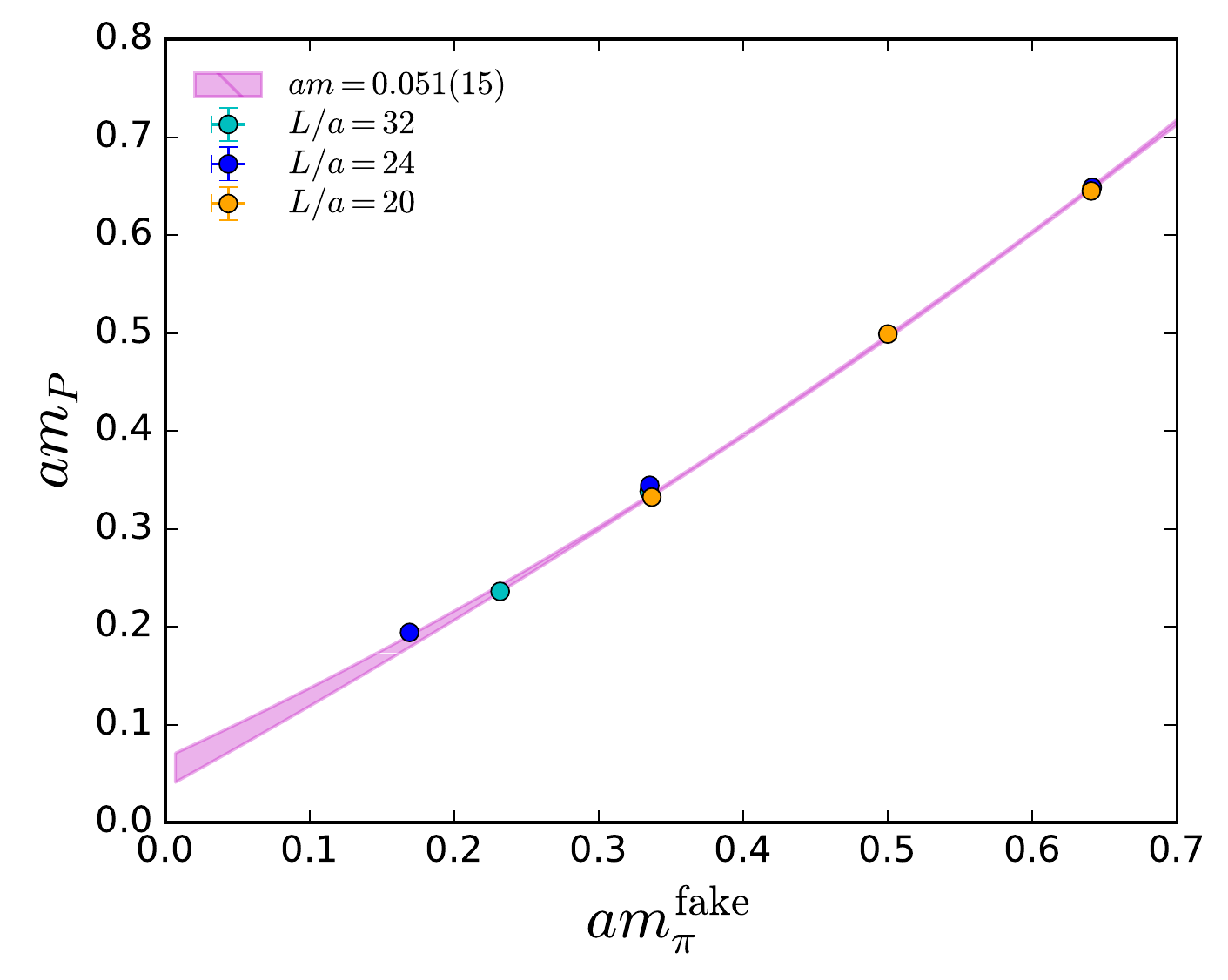}
}
\subfloat[Vector]{
  \includegraphics[width=0.49\textwidth]{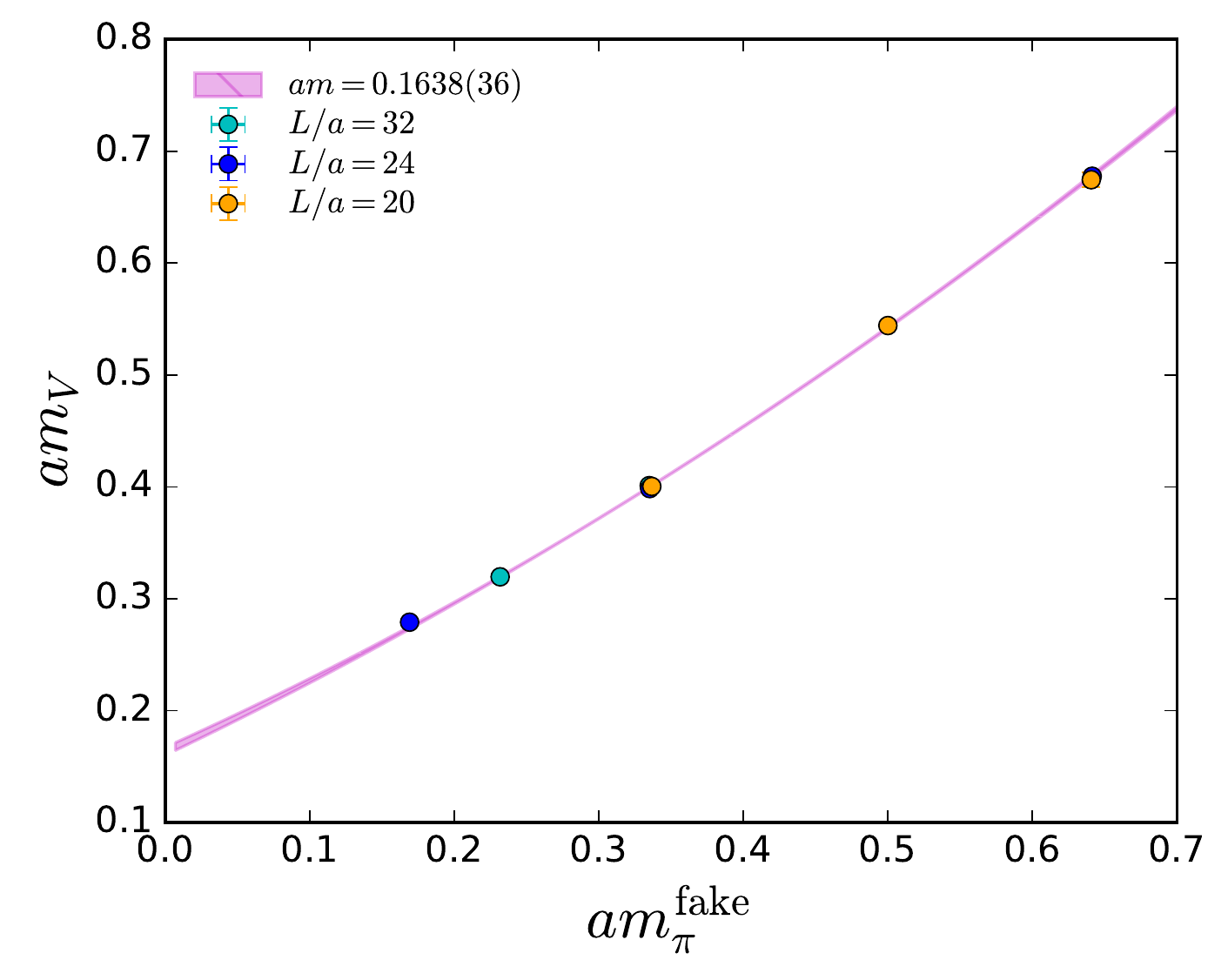}
}
\hspace{0mm}
\subfloat[Scalar]{
  \includegraphics[width=0.49\textwidth]{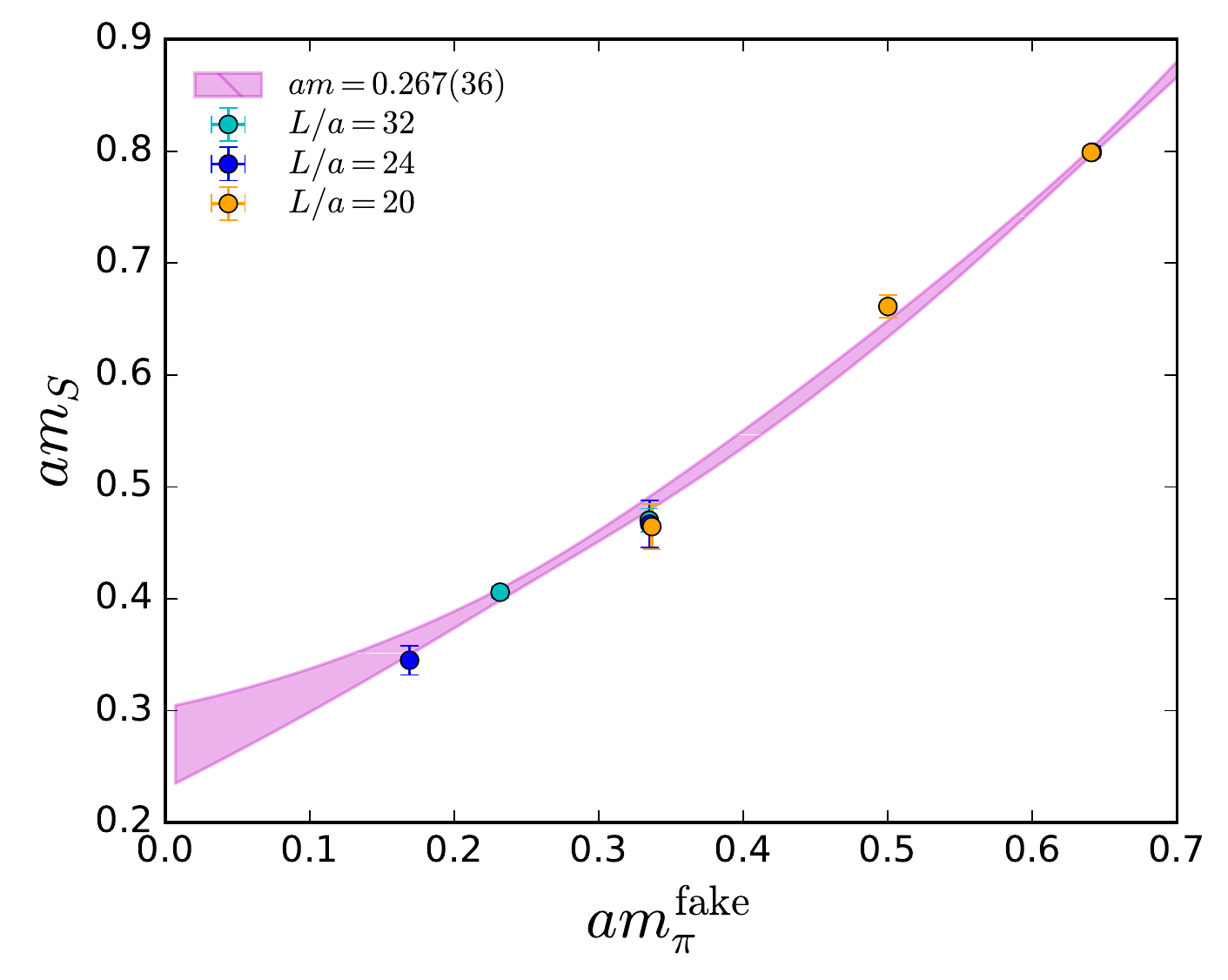}
}
\caption{Examples of the chiral extrapolation of the pseudo-scalar (top left), vector (top right) and scalar-glue (bottom) meson masses to vanishing quark masses.}
\label{fig.extra}
\end{figure}

Some of the extracted energies strongly depend on the volume, whereas others display a clear quark mass dependence. This behaviour is most visible for the scalar-glue correlation function, which is shown in Figure~\ref{fig.scalar}. By carefully studying the quark-mass and volume dependencies of the meson masses and overlap coefficients we disentangle mass-independent (glueballs and torelons) from mesonic states. One conclusion from this study is  that the smaller volumes are affected by sizeable finite-volume effects, especially for the smallest quark masses used here. For the chiral extrapolations, we explore linear and quadratic forms in terms of the fake pion mass with and without an extrapolation to infinite volume. Examples of these chiral extrapolation fits for the pseudo-scalar, vector, and scalar mesons are shown in Figure~\ref{fig.extra}. We note that in the chiral limit the mass of the pseudo-scalar meson extrapolates to a non-vanishing value. This is in agreement with expectations due to the disconnected contributions.

\section{One Flavour for $N_c > 3$}
In order to simulate larger numbers of colours on the lattice, we changed our simulation software to \texttt{Grid} \cite{Boyle:2015tjk}. As an initial test, we explore quenched calculations, for which we employ the Symanzik gauge action~\cite{Iwasaki:1983iya}. The large-$N_c$ limit in pure gauge theory was proposed in the seminal work by t'Hooft \cite{tHooft:1973alw}. Accordingly, we rescale $\beta\propto N_c^2$. Our choices are listed in table~\ref{tab.larger_nc_runs}. \par
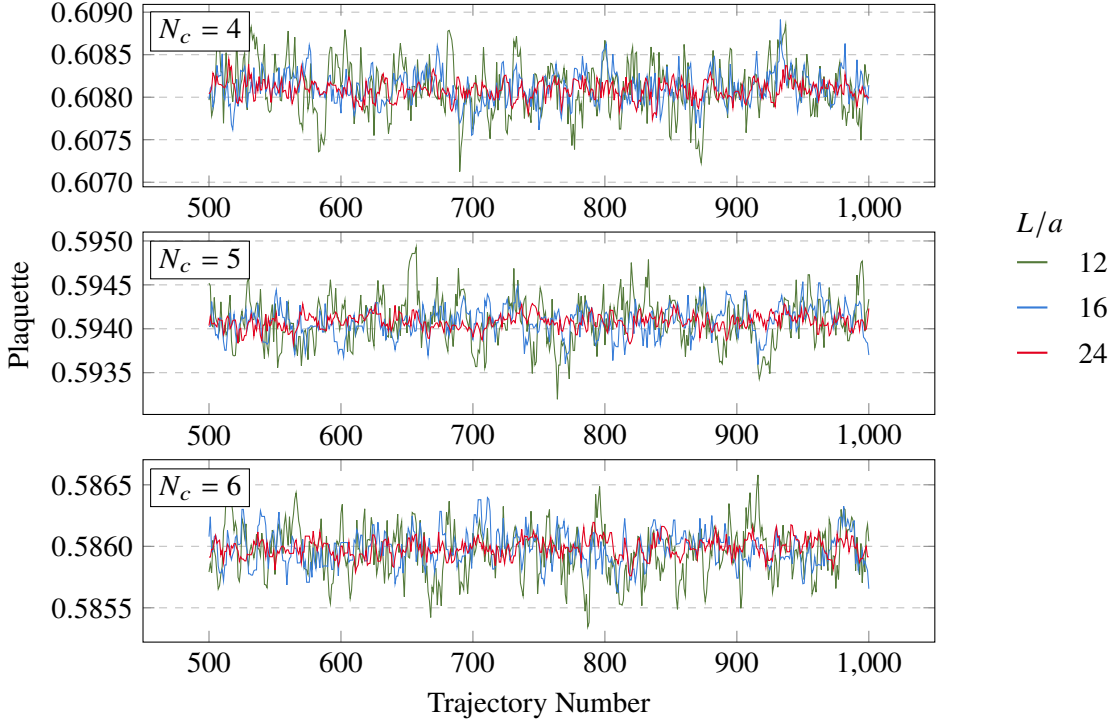
\begin{figure}
\centering
\begin{tikzpicture}
  \begin{axis}[ name=nc4,
                height=4cm,
                width=12cm,
                ymajorgrids=true,
                grid style=dashed, 
                yticklabel style={
                    /pgf/number format/fixed,
                    /pgf/number format/precision=4,
                    /pgf/number format/zerofill
                }]
    \addplot[mark=none,color=mygreen] table[x=x, y=y] {data/plaq_nc4L12.csv};
    \addplot[mark=none,color=myblue] table[x=x, y=y] {data/plaq_nc4L16.csv};
    \addplot[mark=none,color=myred] table[x=x, y=y] {data/plaq_nc4L24.csv};
  \end{axis}
  \begin{axis}[ name=nc5,
                at={($(nc4.south) - (0,6mm)$)},
                anchor=north,
                height=4cm,
                width=12cm, 
                ymajorgrids=true,
                grid style=dashed, 
                yticklabel style={
                    /pgf/number format/fixed,
                    /pgf/number format/precision=4,
                    /pgf/number format/zerofill
                }]
        \addplot[mark=none,color=mygreen] table[x=x, y=y] {data/plaq_nc5L12.csv};
        \addplot[mark=none,color=myblue] table[x=x, y=y] {data/plaq_nc5L16.csv};
        \addplot[mark=none,color=myred] table[x=x, y=y] {data/plaq_nc5L24.csv};
  \end{axis}
  \begin{axis}[ name=nc6,
                at={($(nc5.south)-(0,6mm)$)}, 
                anchor=north,
                height=4cm,
                width=12cm, 
                ymajorgrids=true,
                grid style=dashed, 
                xlabel=Trajectory Number, 
                yticklabel style={
                    /pgf/number format/fixed,
                    /pgf/number format/precision=4,
                    /pgf/number format/zerofill
                }]
        \addplot[mark=none,color=mygreen] table[x=x, y=y] {data/plaq_nc6L12.csv};
        \addplot[mark=none,color=myblue] table[x=x, y=y] {data/plaq_nc6L16.csv};
        \addplot[mark=none,color=myred] table[x=x, y=y] {data/plaq_nc6L24.csv};
  \end{axis}
  \node at (11.8, -0.5) {$L/a$};
  \draw[color=mygreen,thick] (11.5,-1) -- (11.9,-1);
  \node at (12.5,-1) {12};
  \draw[color=myblue,thick] (11.5,-1.6) -- (11.9,-1.6);
  \node at (12.5,-1.6) {16};
  \draw[color=myred, thick] (11.5,-2.2) -- (11.9,-2.2);
  \node at (12.5,-2.2) {24};
  
  \draw[fill=white] (0.1,2.32) -- (1.4,2.32) -- (1.4,1.8) -- (0.1,1.8) -- (0.1,2.32);
  \node at (0.75,2.07) {$N_c=4$};

  \draw[fill=white, yshift=-3.02cm] (0.1,2.3) -- (1.4,2.3) -- (1.4,1.8) -- (0.1,1.8) -- (0.1,2.3);
  \node[yshift=-3.02cm] at (0.75,2.05) {$N_c=5$};

  \draw[fill=white, yshift=-6.04cm] (0.1,2.3) -- (1.4,2.3) -- (1.4,1.8) -- (0.1,1.8) -- (0.1,2.3);
  \node[yshift=-6.04cm] at (0.75,2.05) {$N_c=6$};
  \node[rotate=90] at (-1.6,-1.7) {Plaquette};
\end{tikzpicture}
\caption{Values of the plaquette as function of MD time for $N_c=4, 5, 6$ (top, middle and bottom panel, respectively) for three different spatial volumes.}
\label{fig.plaqNc4}
\end{figure}

For all setups, we use a trajectory length of 2.0 to minimise auto-correlation effects~\cite{Meyer:2006ty}.
%To maintain an approximately constant string tension, it is necessary to scale $\beta$ depending on the number of colours~\cite{Bali:2013kia}. Our choices are listed in table~\ref{tab.larger_nc_runs}. For all setups, we use a trajectory length of 2.0 to minimise auto-correlations~\cite{Meyer:2006ty}.
\begin{table}
\centering
    \begin{tabular}{lrrrrrr}
        \toprule
        Name & $N_c$ & $L/a$ & $\beta$ & MD steps & N$_{\mathrm{cnfg}}$ & Acceptance \\
        \midrule
        pgnc4L12&4&12&8.0&30&500 & 97.4\%\\
        pgnc5L12&5&12&12.5&38&500 & 97.0\%\\
        pgnc6L12&6&12&18.0&38&500 & 98.4\%\\
        pgnc4L16&4&16&8.0&30&500 & 95.8\%\\
        pgnc5L16&5&16&12.5&25&500 & 86.4\%\\
        pgnc6L16&6&16&18.0&28&500 & 82.8\%\\
        pgnc4L24&4&24&8.0&30&500 & 93.4\%\\
        pgnc5L24&5&24&12.5&35&500 & 91.4\%\\
        pgnc6L24&6&24&18.0&35&500 & 85.2\%\\
        \bottomrule
    \end{tabular}
    \caption{Overview of the pure gauge simulations for  $N_c=4, 5$ and $6$.}
    \label{tab.larger_nc_runs}
\end{table}
We use a single integrator level with step sizes tuned for acceptance above 80\%. We observe that the 2nd-order Omelyan integrator implemented in \texttt{Grid} scales unsatisfactorily with the numbers of colours and therefore use a force-gradient integrator \cite{Chin:2000zz}. The Hamiltonian violations of this integrator type are smaller and are less impacted by increases in the lattice volume, see \cite{Kennedy:2012gk}, and number of colours, which lead to a significant performance improvement in \texttt{Grid} for $N_c>3$ in our preliminary examinations.
Figure~\ref{fig.plaqNc4} shows the plaquette for three spatial extents ($L/a=12, 16, 24$) and fixed time-extent of $T/a=48$ after a conservative estimate of 500 thermalisation trajectories.

\begin{figure}
\centering
\begin{tikzpicture}
  \begin{axis}[ name=nc4,
                height=4cm,
                width=9cm,
                ymajorgrids=true,
                grid style=dashed, 
                ymin=-12.5, ymax=12.5]
    \addplot[mark=none,color=mygreen] table[x=x, y=y] {data/tc_nc4L12.csv};
    \addplot[mark=none,color=myblue] table[x=x, y=y] {data/tc_nc4L16.csv};
    \addplot[mark=none,color=myred] table[x=x, y=y] {data/tc_nc4L24.csv};
  \end{axis}
  \begin{axis}[ name=hist_nc4, 
                at={($(nc4.east) + (3cm,0)$)},
                anchor=south,
                rotate around={-90:(current axis.origin)},
                x dir=reverse,
                height=4cm, 
                width=4cm, 
                ymajorgrids=true, 
                grid style=dashed, 
                ybar interval, 
                xmin=-12.5, xmax=12.5, 
                ymin=0, ymax=90,
                xtick={-10,0,10},
                xticklabel=\empty, 
                yticklabel=\empty]
    \addplot[mark=none,color=mygreen,fill=mygreen] table[x=breaks, y=counts]{data/tc_hist_nc4L12.csv};
    \addplot[mark=none,color=myblue,fill=myblue] table[x=breaks, y=counts]{data/tc_hist_nc4L16.csv};
    \addplot[mark=none,color=myred,fill=myred] table[x=breaks, y=counts]{data/tc_hist_nc4L24.csv};
  \end{axis}             
  \begin{axis}[ name=nc5,
                at={($(nc4.south) - (0,6mm)$)},
                anchor=north,
                height=4cm,
                width=9cm, 
                ymajorgrids=true,
                grid style=dashed, 
                ylabel=Topological Charge, 
                ymin=-12.5, ymax=12.5]
    \addplot[mark=none,color=mygreen] table[x=x, y=y] {data/tc_nc5L12.csv};
    \addplot[mark=none,color=myblue] table[x=x, y=y] {data/tc_nc5L16.csv};
    \addplot[mark=none,color=myred] table[x=x, y=y] {data/tc_nc5L24.csv};
  \end{axis}
  \begin{axis}[ name=hist_nc5, 
                at={($(nc5.east) + (3cm,0)$)},
                anchor=south,
                rotate around={-90:(current axis.origin)},
                x dir=reverse,
                height=4cm, 
                width=4cm, 
                ymajorgrids=true, 
                grid style=dashed, 
                ybar interval,
                xmin=-12.5, xmax=12.5, 
                ymin=0, ymax=90,
                xtick={-10,0,10},
                xticklabel=\empty, 
                yticklabel=\empty]
    \addplot[color=mygreen,fill=mygreen] table[x=breaks, y=counts]{data/tc_hist_nc5L12.csv};
    \addplot[color=myblue,fill=myblue] table[x=breaks, y=counts]{data/tc_hist_nc5L16.csv};
    \addplot[color=myred,fill=myred] table[x=breaks, y=counts]{data/tc_hist_nc5L24.csv};
  \end{axis} 
  \begin{axis}[ name=nc6,
                at={($(nc5.south)-(0,6mm)$)}, 
                anchor=north,
                height=4cm,
                width=9cm, 
                ymajorgrids=true,
                grid style=dashed, 
                xlabel=Trajectory Number,
                ymin=-12.5, ymax=12.5]
    \addplot[mark=none,color=mygreen] table[x=x, y=y] {data/tc_nc6L12.csv};
    \addplot[mark=none,color=myblue] table[x=x, y=y] {data/tc_nc6L16.csv};
    \addplot[mark=none,color=myred] table[x=x, y=y] {data/tc_nc6L24.csv};
  \end{axis}
  \begin{axis}[ name=hist_nc6, 
                at={($(nc6.east) + (3cm,0)$)},
                anchor=south,
                rotate around={-90:(current axis.origin)},
                x dir=reverse,
                height=4cm, 
                width=4cm, 
                ymajorgrids=true, 
                grid style=dashed, 
                ybar interval,
                xmin=-12.5, xmax=12.5,
                ymin=0, ymax=90,
                xticklabel=\empty, 
                yticklabel=\empty,
                yticklabel pos=right,
                xtick={-10,0,10}]
    \addplot[mark=none,color=mygreen,fill=mygreen] table[x=breaks, y=counts]{data/tc_hist_nc6L12.csv};
    \addplot[mark=none,color=myblue,fill=myblue] table[x=breaks, y=counts]{data/tc_hist_nc6L16.csv};
    \addplot[mark=none,color=myred,fill=myred] table[x=breaks, y=counts]{data/tc_hist_nc6L24.csv};
  \end{axis} 
  \node at (11.8, -0.5) {$L/a$};
  \draw[color=mygreen,thick] (11.5,-1) -- (11.9,-1);
  \node at (12.5,-1) {12};
  \draw[color=myblue,thick] (11.5,-1.6) -- (11.9,-1.6);
  \node at (12.5,-1.6) {16};
  \draw[color=myred, thick] (11.5,-2.2) -- (11.9,-2.2);
  \node at (12.5,-2.2) {24};
  
  \draw[fill=white] (0.1,2.32) -- (1.4,2.32) -- (1.4,1.8) -- (0.1,1.8) -- (0.1,2.32);
  \node at (0.75,2.07) {$N_c=4$};

  \draw[fill=white, yshift=-3.02cm] (0.1,2.3) -- (1.4,2.3) -- (1.4,1.8) -- (0.1,1.8) -- (0.1,2.3);
  \node[yshift=-3.02cm] at (0.75,2.05) {$N_c=5$};

  \draw[fill=white, yshift=-6.04cm] (0.1,2.3) -- (1.4,2.3) -- (1.4,1.8) -- (0.1,1.8) -- (0.1,2.3);
  \node[yshift=-6.04cm] at (0.75,2.05) {$N_c=6$};

  \node at (8,-6.3) {0};
  \node at (9.1,-6.3) {40};
  \node at (10.2,-6.3) {80};
  \node at (9.2, -6.85) {Frequency};
\end{tikzpicture}
\caption{Topological charge for different number of colours and spatial volumes.}
\label{fig.tc}
\end{figure}
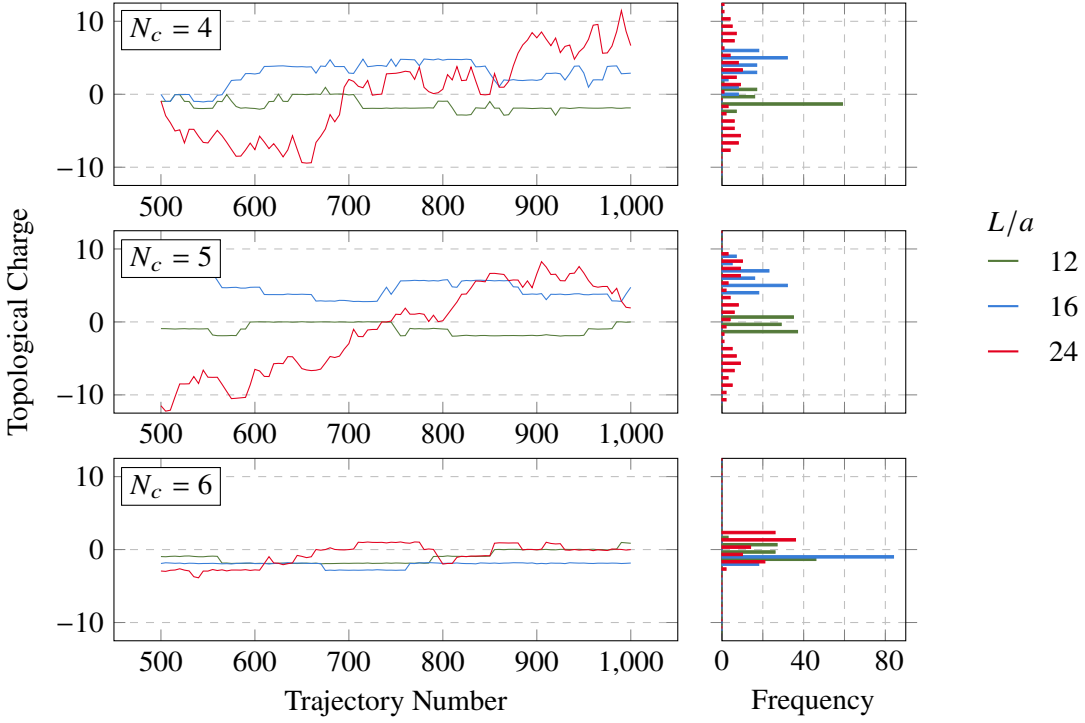

The evolution of the topological charge is shown in Figure~\ref{fig.tc}. As expected, we observe that fluctuations in the topological charge increase as the volume increases. However, histograms in the right column of this Figure show that we have not yet generated sufficiently many configurations to draw clear conclusions about the exploration of different topological regimes. We observe indications that topological freezing is exacerbated at larger $N_c$, due to the need to simulate at values of $\beta$ as large as $18$.

\section{Conclusions and Outlook}
We have presented an update on our current efforts to verify the predictions from supersymmetric Yang-Mills theories by studying the mesonic spectrum of QCD with one flavour at $N_c=3$. Additionally, we have shown that extending our simulations to larger $N_c$ becomes more costly, in particular as the topological charge shows significantly reduced fluctuations.  

\section*{Acknowledgements}
This project has received funding from the European Union's Horizon 2020 research and innovation programme under the Marie Sk\l odowska-Curie grant agreement \textnumero~813942 and 894103. M.D.M. and J.T.T. are partially supported by DFF Research project 1. Grant n. 8021-00122B. The computations were performed on UCloud interactive HPC system and ABACUS2.0 supercomputer, which are managed by the eScience Center at the University of Southern Denmark and LUMI supercomputer, hosted by the LUMI consortium, as part of the LUMI-G pilot program.

\bibliographystyle{apsrev4-1}
\bibliography{oneflavour}

\end{document}